\documentclass[aps,showpacs,twocolumn]{revtex4}
\usepackage{epsfig}

\begin{document}

\title{Baryonia and near-threshold enhancements}

\author{Chengrong Deng$^1$, Jialun Ping$^2$\footnote{Corresponding
 author}\footnote{Electronic address:
jlping@njnu.edu.cn}, Youchang Yang$^3$, and Fan Wang$^4$}

\affiliation{$^1$School of Mathematics and Physics, Chongqing
Jiaotong University, Chongqing 400074, P.R. China}

\affiliation{$^2$Department of Physics, Nanjing Normal University,
Nanjing 210097, P.R. China}

\affiliation{$^3$Department of Physics, Zunyi Normal College,
Zunyi 563002, P.R. China}

\affiliation{$^4$Department of Physics, Nanjing University,
Nanjing 210093, P.R. China}

\begin{abstract}
The baryon-antibaryon spectrum consisting of strange, charm and
bottom quarks is studied in the color flux-tube model with a
multi-body confinement interaction. Numerical results indicate
that many low-spin baryon-antibaryon states can form compact
hexaquark states and are stable against the decay into a baryon
and an antibaryon. The multi-body confinement interaction as a
binding mechanism plays an important role in the formation of
the states. They can be searched in the $e^+e^-$ annihilation and
charmonium or bottomonium decay if they really exist. The
newly reported states, $X(1835)$,
$X(2370)$, $Y(2175)$, $Y(4360)$ and $Y_b(10890)$, may be
interpreted as $N\bar{N}$, $\Delta\bar{\Delta}$,
$\Lambda\bar{\Lambda}$, $\Lambda_c\bar{\Lambda}_c$ and
$\Lambda_b\bar{\Lambda}_b$ states, respectively.
\end{abstract}

\pacs{12.39.Jh, 13.75.Cs, 14.40.Rt}

\maketitle

\section{introduction}

The research of baryonia  has a rather long history, which can
date back to 1940s. Fermi and Yang
proposed that the $\pi$-meson may be a composite particle of
nucleon-antinucleon ($N\bar{N}$)~\cite{fermi}, in which a strong attractive force is
assumed to bind them together because of the mass of $\pi$-meson
is substantially smaller than twice the mass of a nucleon.
Subsequently, Sakata extended Fermi and Yang's idea by introducing
a strange baryon $\Lambda$ and its antiparticle. The strange
baryon $\Lambda$, proton $p$ and neutron $n$ and their
antiparticles were regarded as the fundamental building blocks to construct other
mesons and baryons, which is the well known Fermi-Yang-Sakata
(FYS) model~\cite{sakata}. The profound difficulty of the FYS
model was the enormous binding energy for sticking a baryon and an
antibaryon together to form a light meson, the FYS model was
therefore replaced by quark models. Many researchers abandoned the
FYS's point of view and pioneered that $N\bar{N}$ states were no
more associated with ``ordinary" light mesons, but instead with
new types of mesons with a mass near the $N\bar{N}$ threshold and
specific decay properties~\cite{baryonia}.

In recent years, many near-threshold enhancements are observed in
experiments, the $p\bar{p}$ enhancement is observed in the
$J/\psi\rightarrow \gamma p\bar{p}$,
$\psi^{\prime}\rightarrow\pi^0p\bar{p}$ $\eta p\bar{p}$, and
$B^{\pm}\rightarrow p\bar{p}K^{\pm}$~\cite{ppbar-exp}; the
$\Lambda\bar{\Lambda}$ enhancement is observed in the
$B^{+}\rightarrow \Lambda\bar{\Lambda}K^{+}$ and
$e^+e^-\rightarrow \Lambda\bar{\Lambda}$~\cite{lambda}; the
$\Lambda_c^+\Lambda_c^-$ enhancement is observed in the
$e^+e^-\rightarrow \Lambda_c^+\Lambda_c^-$~\cite{lambdac}, et al.
Furthermore, many other resonances called $XYZ$ particles were
also observed in experiments. It is hard to accommodate some of
them, such as $X(3872)$ and $Y(4260)$, into quark models due to
their extraordinary properties, which goes beyond our anticipation,
because it is taken for granted that the heavy mesons can be well
described with quark models. The appearance of XYZ particles
forces us to propose various interpretations rather than a
$q\bar{q}$ configuration to clarify the structure of
the states. The recent progresses and a rather complete list of
references on XYZ particles can be found in recent review and
references therein~\cite{review}. In addition to the explanations
of theses exotic states as tetraquark states, meson-meson molecular states,
hybrid quarkonia and orbital excited states of conventional mesons, et
al, baryonia or hexaquark states $q^3\bar{q}^3$ are also possible
interpretations. The
$p\bar{p}$ enhancement was interpreted as a baryonium $N\bar{N}$
with quantum numbers $J^{PC}=0^{-+}$ in the different theoretical
frames by many authors~\cite{ppbar-the}. The states $Y(4260)$,
$Y(4361)$, $Z^{\pm}(4430)$ and $Y(4664)$ were systematically
embedded into an extended baryonium picture~\cite{cfqiao}. The
$Y(2175)$ was described as a bound state $\Lambda\bar{\Lambda}$
with quantum numbers $^{2S+1}L_J$=$^3S_1$ in the
one-boson-exchange potential model~\cite{lzhao}.

Nonstrange hexaquark states $q^3\bar{q}^3$ were systematically
studied in a color flux-tube model with a six-body confinement
potential instead of an additive two-body interaction in our
previous work, and it was found that some ground states are stable
against disintegrating into a baryon and an
anti-baryon~\cite{deng1}. In the present work, the spectrum of
the baryonia of the ground state consisting of strange,
charm and bottom quarks is studied in the color flux-tube model. The
work is not only a natural extension of the previous work to try
to interpret the structure of the recently discovered resonances,
but also provides a new insight into exploring the baryonium states.
The research shows that many low-spin baryon-antibaryon states are
compact hexaquark states and stable against direct decaying into a
baryon and an antibaryon, a multi-body confinement interaction in
the model plays an important role in the short-range domain. The
dominant components of the new hadron states $X(1835)$, $X(2370)$,
$Y(2175)$, $Y(4260)$ and $Y_b(10890)$ may be interpreted as
$N\bar{N}$, $\Delta\bar{\Delta}$, $\Lambda\bar{\Lambda}$,
$\Lambda_c\bar{\Lambda}_c$ and $\Lambda_b\bar{\Lambda}_b$ bound
states, respectively.

The paper is organized as follows: Section II is devoted to the
descriptions of the color flux-tube model and gives the
Hamiltonian of baryons and hexaquark systems. A brief
introductions of the constructions of the wave functions of
baryons and hexaquark systems are given in Sec. III. The numerical
results and discussions are presented in Sec. IV. A brief summary
is given in the last section.

\section{color flux-tube model and hamiltonian}

Quantum Chromodynamics (QCD) is widely accepted as the fundamental
theory to describe the strong interacting systems and has
verified in high momentum transfer process. In the low energy
region, such as hadron spectroscopy and hadron-hadron interaction
study, the \emph{ab initio} calculation directly from QCD becomes
very difficult due to the complication of nonperturbative nature.
Although many nonperturbative methods have been developed, such as
lattice QCD (LQCD), QCD sum rule, large-$N_c$ expansion, chiral
unitary theory, et al, the QCD-inspired constituent quark model (CQM)
is still a useful tool in obtaining physical insight for these
complicated strong interacting systems. CQM can offer the most
complete description of hadron properties and is probably the most
successful phenomenological model of hadron
structure~\cite{godfrey}.

CQM is formulated under the assumption that the hadrons are color
singlet non-relativistic bound states of constituent quarks with
phenomenological effective masses and interactions. The effective
interactions includes one gluon exchange (OGE), one boson exchange
(OBE) and a confinement potential. Traditional CQM includes the
typical Isgur-Karl model and chiral quark
model~\cite{isgurkarl,chiral}, in which the confinement potential
can be phenomenologically described as the sum of two-body
interactions proportional to the color charges and $r_{ij}^k$,
\begin{eqnarray}
V^C&=& -a_c\sum_{i>j}^n\mathbf{\lambda}_{i}\cdot\mathbf{\lambda}_{j}r^k_{ij}
\end{eqnarray}
where $r_{ij}$ is the distance between two interacting quarks and $k$
usually takes 1 or 2. The traditional models can describe the
properties of ordinary hadrons ($q^3$ and $q\bar{q}$) well. However,
the traditional models lead to power law van der Waals forces
between color-singlet hadrons and the anti-confinement in a color
symmetrical quark or antiquark pair. The problems are related to
the fact that the traditional CQM does not respect local color
gauge invariance.

The color flux-tube structures of ordinary hadrons are unique and
trivial, many important low-energy QCD information may be absent
in the descriptions of these objects, such as quark pair in color
symmetrical $\mathbf{6}$ ($\bar{\mathbf{6}}$) representation. Multiquark
systems, if they really exist, have various color flux-tube
structures in the intermediate- and short-distance domains, which
may contain abundant low-energy QCD information and affect the
properties of multiquark
systems~\cite{scalar,cyclobutadiene,deng2,ping,ww}. The mixing
effect of the color flux-tube structures can provide the
intermediate-range attractive force coming from the $\sigma$ meson
or $\pi\pi$ exchange~\cite{hxhuang}. The traditional CQM is hard
to describe various color flux-tube structures of multiquark
systems.

LQCD calculations of ordinary hadrons, tetraquark and pentaquark
states reveal various color flux-tube structures~\cite{lattice}.
Within the color flux-tube picture, the confinement potential of
multiquark states is a multibody interaction and can be simulated
by a potential which proportional to the minimum of the total
length of all color flux tubes which connects the quarks
(antiquarks) to form a multiquark system~\cite{lattice}. Based on
the traditional CQM and the LQCD picture, the color flux-tube
model has been developed to study multiquark systems, in which a
multibody confinement interaction is employed, and a sum of the
square of the length of flux tubes rather than a linear one is
assumed to simplify the
calculation~\cite{scalar,cyclobutadiene,deng2,ping,ww}. The
approximation is justified because of the following two reasons:
one is that the spatial variations in separation of the quarks
(lengths of the flux tube) in different hadrons do not differ
significantly, so the difference between the two functional forms
is small and can be absorbed in the adjustable parameter, the
stiffness. The other is that we are using a nonrelativistic
dynamics in the study. As was shown long ago~\cite{goldman}, an
interaction energy that varies linearly with separation between
fermions in a relativistic first order differential dynamics has a
wide region in which a harmonic approximation is valid for the
second order (Feynman-Gell-Mann) reduction of the equations of
motion. The comparative studies also indicated that the difference
between the two type confinement potentials is very
small~\cite{ping,ww}.

The description of the properties of ordinary hadrons is the
starting point of the phenomenological investigation of multiquark
systems. The Y-shaped color flux-tube structure, the LQCD picture
of a baryon~\cite{bissey}, is shown in Fig.1, in which
$\mathbf{r}_i$ represents the spatial position of the $i$-th quark
denoted by a black dot and $\mathbf{y}_0$ denotes a junction where
three color flux tubes meet. In the color flux-tube model with
quadratic confinement potential, the three-body potential can be
written as
\begin{eqnarray}
V^C(3)=K\left((\mathbf{r}_1-\mathbf{y}_0)^2+(\mathbf{r}_2-\mathbf{y}_0)^2
+(\mathbf{r}_3-\mathbf{y}_0)^2\right)
\end{eqnarray}
the position of the junction $\mathbf{y}_0$ can be fixed by
minimizing the energy of baryons, then we get
\begin{equation}
\mathbf{y}_0=\frac{\mathbf{r}_1+\mathbf{r}_2+\mathbf{r}_3}{3}
\end{equation}
the minimum of the confinement potential for baryons has therefore
the following forms
\begin{equation}
V_{min}^{C}(3)=
K\left(\left(\frac{\mathbf{r}_1-\mathbf{r}_2}{\sqrt{2}}\right)^2
+\left(\frac{2\mathbf{r}_3-\mathbf{r}_1-\mathbf{r}_2}{\sqrt{6}}\right)^2\right)
\end{equation}
the above equation can also be expressed as the sum of three pairs
of two-body interactions,
\begin{equation}
V_{min}^{C}(3)=\frac{K}{3}\left(({\mathbf{r}_1-\mathbf{r}_2})^2
+({\mathbf{r}_2-\mathbf{r}_3})^2+({\mathbf{r}_1-\mathbf{r}_3})^2\right)
\end{equation}
It can be seen that the three-body quadratic confinement potential
of a baryon is totally equivalent to the sum of the two-body one,
see the $\Delta$-shaped structure in Fig.1, although the equivalence is
only approximately valid for the linear confinement potential.
\begin{figure}[ht]
\epsfxsize=2.5in \epsfbox{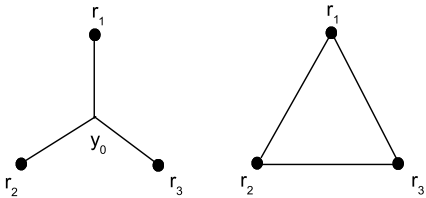} \caption{Three-body (left) and
two-body (right) confinement potential}
\end{figure}

With respect to a hexaquark system $q^3\bar{q}^3$, four possible
color flux-tube structures are listed in Fig.2, in which a black
dot denotes a quark and a hollow dot denotes an antiquark. The
structure (a) is a baryon-antibaryon molecule state:
$[q^3]_{1}[\bar{q}^3]_{1}$; The hexaquark state with the structure
(b) is called a color octet baryon-antibaryon state:
$\left [[q^3]_{8}[\bar{q}^3]_{8}\right ]_1$. In the case of the
structure (c), it can be called a diquark-antidiquark
hexaquark state: $\left[\left[[q^2]_{\bar{3}}\bar{q}\right]_3
\left[[\bar{q}^2]_{3}q\right]_{\bar{3}}\right]_1$. The last one is
similar to a chemical benzene, and it is therefore called QCD
benzene, this structure for a hexaquark state $q^6$ was studied in
our previous work~\cite{ping}. Of cause, the color flux-tube
structures should include three color singlet mesons
configuration: $[q\bar{q}]_1[q\bar{q}]_1[q\bar{q}]_1$, which  must
be taken into account in the decay of a $q^3\bar{q}^3$ system into
three mesons, this task is left as our future work. The color
flux-tube in hadrons should be very similar to the chemical bond
in organic compounds. The same molecular constituents may have
different chemical bond structure; those are called isomeric
compounds. Therefore, the multiquark systems with the same quark
content but different flux-tube structures are similarly called
QCD isomeric compounds.

In general, a hexaquark system should be the mixture of all
possible flux-tube structures. In order to avoid a too complicated
calculation in the present work, only the first two structures in
Fig.2 are considered. Within the color flux-tube model, the
confinement potential for the structure (a) can be written
as
\begin{eqnarray}
V_{min}^a(6)&=&
K\left(\left(\frac{\mathbf{r}_1-\mathbf{r}_2}{\sqrt{2}}\right)^2
+\left(\frac{2\mathbf{r}_3-\mathbf{r}_1-\mathbf{r}_2}{\sqrt{6}}\right)^2
\right. \nonumber \\
&+&
\left.\left(\frac{\mathbf{r}_4-\mathbf{r}_5}{\sqrt{2}}\right)^2
+\left(\frac{2\mathbf{r}_6-\mathbf{r}_4-\mathbf{r}_5}{\sqrt{6}}\right)^2
\right)
\end{eqnarray}
while the confinement potential for the structure (b) has the following form
\begin{eqnarray}
V^{b}(6)&=&K\left((\mathbf{r}_1-\mathbf{y}_1)^2+(\mathbf{r}_2-\mathbf{y}_1)^2
+(\mathbf{r}_3-\mathbf{y}_2)^2\right. \nonumber \\
&+&
\left.(\mathbf{r}_6-\mathbf{y}_3)^2+(\mathbf{r}_4-\mathbf{y}_4)^2
+(\mathbf{r}_5-\mathbf{y}_4)^2 \right. \nonumber\\
&+&
\left.\kappa_{d_{12}}(\mathbf{y}_1-\mathbf{y}_2)^2+\kappa_{d_{23}}(\mathbf{y}_2
-\mathbf{y}_3)^2 \right. \nonumber\\
&+&\left.\kappa_{d_{34}}(\mathbf{y}_3-\mathbf{y}_4)^2 \right)
\end{eqnarray}

\begin{figure}
 \epsfxsize=3.9in \epsfbox{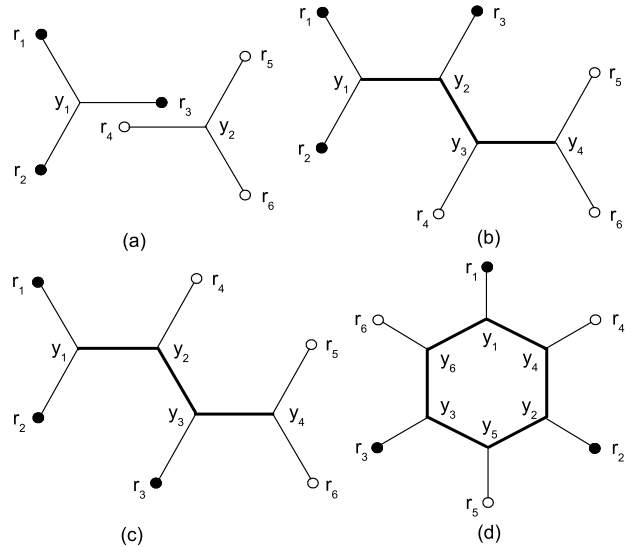}
\caption{Four color flux-tube structures of a $q^3\bar{q}^3$
system.}
\end{figure}

In above equation, $K$ is the stiffness constant of an elementary
or color triplet flux-tube, while $K\kappa_{d_{ij}}$ is other
color flux-tube stiffness called as compound ones. The compound
color flux-tube stiffness parameter $\kappa_{d_{ij}}$ depends on
the color dimension, $d_{ij}$, of the string,
\begin{equation}
 \kappa_{d_{ij}}=\frac{C_{d_{ij}}}{C_3},
\end{equation}
where $C_{d_{ij}}$ is the eigenvalue of the Casimir operator
associated with the $SU(3)$ color representation $d_{ij}$ on
either end of the string, namely $C_3=\frac{4}{3}$,
$C_6=\frac{10}{3}$ and $C_8=3$. For the sake of simplicity, the
average $\kappa_d$ for $\kappa_{d_{ij}}$ is used in numerical
calculations.

For given quark (antiquark) positions $\mathbf{r}_i$, those
junctions $\mathbf{y}_i$ are obtained by minimizing the
confinement potential. By introducing the following set of
canonical coordinates $\mathbf{R}_i$,
\begin{eqnarray}
\mathbf{R}_1&=&\frac{1}{\sqrt{2}}(\mathbf{r}_1-\mathbf{r}_2),
~~\mathbf{R}_2=\frac{1}{\sqrt{2}}(\mathbf{r}_4-\mathbf{r}_5)\nonumber\\
\mathbf{R}_3&=&\frac{1}{\sqrt{12}}(\mathbf{r}_1+\mathbf{r}_2
-2\mathbf{r}_3+\mathbf{r}_4+\mathbf{r}_5-2\mathbf{r}_6)\nonumber\\
\mathbf{R}_4&=&\frac{1}{\sqrt{33+5\sqrt{33}}}(\mathbf{r}_1+\mathbf{r}_2
-w_1\mathbf{r}_3-\mathbf{r}_4-\mathbf{r}_5+w_1\mathbf{r}_6)\nonumber\\
\mathbf{R}_5&=&\frac{1}{\sqrt{33-5\sqrt{33}}}\left(\mathbf{r}_1+\mathbf{r}_2
+w_2\mathbf{r}_3-\mathbf{r}_4-\mathbf{r}_5-w_2\mathbf{r}_6\right)\nonumber\\
\mathbf{R}_6&=&\frac{1}{\sqrt{6}}(\mathbf{r}_1+\mathbf{r}_2+\mathbf{r}_3
+\mathbf{r}_4+\mathbf{r}_5+\mathbf{r}_6).
\end{eqnarray}
the minimum of the confinement potential takes the following form,
\begin{eqnarray}
V^{b}_{min}(6)&=&K\left(\mathbf{R}_1^2+\mathbf{R}_2^2+\frac{3\kappa_d}{2+3\kappa_d}\mathbf{R}_3^2
\right.
\\&+&\left.\frac{2\kappa_d(\kappa_d+w_3)}{2\kappa_d^2+7\kappa_d+2}\mathbf{R}_4^2
+\frac{2\kappa_d(\kappa_d+w_4)}{2\kappa_d^2+7\kappa_d+2}\mathbf{R}_5^2\right)\nonumber
\end{eqnarray}
where $w_1=\frac{\sqrt{33}+5}{2}$, $w_2=\frac{\sqrt{33}-5}{2}$,
$w_3=\frac{7+\sqrt{33}}{4}$, and $w_4=\frac{7-\sqrt{33}}{4}$.
Clearly this confinement potential is a multibody interaction
rather than the sum of two-body one.
% in the sense that a move of a quark may affect flux tubes connecting pattern.
When two clusters
$q^3$ and $\bar{q}^3$ separate largely, a baryon and an antibaryon
should be a dominant component of the system because other hidden
color flux-tube structures are suppressed due to the color confinement.
With the separation reducing, a hadronic molecule state may be
formed if the attractive force between a baryon and an antibaryon
is strong enough. When they are close enough to be within the
range of confinement (about 1 fm), all possible flux-tube
structures may appear due to the excitation and rearrangement of
color flux tubes. In this case, the confinement potential of the
system should at least be taken to be the minimum of the two
flux-tube structures. It therefore reads
\begin{eqnarray}
V_{min}^{C}(6) = \mbox{min} \left( V^{a}_{min}, V^{b}_{min}
\right)
\end{eqnarray}

OGE and/or OBE are important and responsible for the mass splitting
in the ordinary hadron spectra. The model with OGE and the model with
OGE+OBE can all describe the ground states of baryons well,
the differences of two models appear in the description of excited
baryons~\cite{klempt}. The study on the ground states of
nonstrange hexaquark systems indicates that the difference of two
models is small~\cite{deng2}, only OGE is therefore taken into
account in the present work. The complete Hamiltonian used here is
listed as the following,
\begin{eqnarray}
H_n & = & \sum_{i=1}^n \left( m_i+\frac{\mathbf{p}_i^2}{2m_i}
\right)-T_{C}+\sum_{i>j}^{n}V_{ij}^G +V_{min}^C(n)\nonumber\\ \\
V_{ij}^G & = & {\frac{1}{4}}\alpha _{s}\mathbf{\lambda}
_{i}\cdot\mathbf{\lambda}_{j}\left({\frac{1}{r_{ij}}}-{\frac{2\pi}{3}}
\delta(\mathbf{r}_{ij}) {\frac{\mathbf{\sigma}_{i}\cdot
\mathbf{\sigma}_{j}}{m_im_j}} \right)
\end{eqnarray}
The tensor forces and spin-orbit forces between quarks are omitted
in the model, because our primary interest is in the lowest
energies and their contributions to the ground states are small or
zero. In the above expression of $H_n$, $n=3$ or $n=6$, $T_{C}$ is the
center-of-mass kinetic energy, $m_i$ and $\mathbf{p}_i$ are the mass and
momentum of the $i$-th quark, $\mathbf{\lambda}$ and $\mathbf{\sigma}$ are
the $SU(3)$ Gell-man and $SU(2)$ Pauli matrices, respectively,
note that $\mathbf{\lambda}\rightarrow-\mathbf{\lambda}^{*}$ for
anti-quark, all other symbols have their usual meanings. An
effective scale-dependent strong coupling constant is used
here~\cite{vijande},
\begin{equation}
\alpha_s(\mu)=\frac{\alpha_0}{\ln\left(\frac{\mu^{2}+\mu_0^2}{\Lambda_0^2}\right)}
\end{equation}
where $\mu$ is the reduced mass of two interactional quarks $q_i$
and $q_j$, namely $\mu=\frac{m_im_j}{m_i+m_j}$, $\Lambda_0$,
$\alpha_0$ and $\mu_0$ are model parameters. The
$\delta$-function, arising as a consequence of the
non-relativistic reduction of the one-gluon exchange diagram
between point-like particles, has to be regularized in order to
perform numerical calculations. It reads~\cite{bhad}
\begin{equation}
\delta(r_{ij})=\frac{1}{\beta^3\pi^{\frac{3}{2}}}e^{-\frac{r^2_{ij}}{\beta^2}}
\end{equation}
where $\beta$ is a model parameter which is determined by fitting
the experiment data.

As far as a baryon is concerned, the color flux-tube model is not
a new one, it reduces to the traditional quark model. However, it being
applied to multiquark systems, the
confinement potential is a multibody interaction instead of a
color dependent two-body one used in traditional quark models
~\cite{scalar,cyclobutadiene,deng2,ping,ww}. In fact, the color
flux-tube model based on traditional quark models and LQCD picture
merely modifies the two-body confinement potential to describe
possible multiquark states with multibody confinement potential.

\section{wave functions and gaussian expansion method}

The total wave function of baryons can be written as the
direct products of color, isospin, spin and spatial
terms,
\begin{eqnarray}
\Phi_{IM_IJM_J}(\mathbf{R}, \mathbf{r})=\chi_c
\left[\Psi^G_{L_TM_T}(\mathbf{R},
\mathbf{r})\eta_{IM_ISM_S}\right]_{IM_IJM_J}
\end{eqnarray}
in which $[\cdots]_{IM_IJM_J}$ means coupling the spin $S$ and
total orbital angular momentum $L_T$ with Clebsch-Gordan
coefficients. The color-part wave function $\chi_c$ is
antisymmetrical because of the color singlet requirement. Only
$u$- and $d$-quark are regarded as identical particles, the
$SU(4)\supset SU_s(2)\times SU_f(2)$ symmetry is therefore used in
the spin-flavor wave function $\eta_{IM_ISM_S}$. The spatial wave
functions of identical particles are assumed to be symmetrical
because we are interesting in the ground states. We can define
Jacobi coordinates $\mathbf{r}_{ij}$ and $\mathbf{R}_k$ for the
cyclic permutations of $(1,2,3)$,
\begin{eqnarray}
\mathbf{r}_{ij}=\mathbf{r}_i-\mathbf{r}_j,\ {} \ {}
\mathbf{R}_k=\mathbf{r}_k-\frac{m_i\mathbf{r}_i+m_j\mathbf{r}_j}{m_i+m_j}
\end{eqnarray}
Within the framework of Gaussian expansion method
(GEM)~\cite{gem}, the total spatial symmetrical wave functions of
baryons with three identical particles, such as $N$, $\Delta$ and
$\Omega$, can be expressed as,
\begin{eqnarray}
\Psi_{L_TM_T}(\mathbf{R}, \mathbf{r})=
{\sum_{i,j,k=1}^{3}}\left[\phi_{lm}(\mathbf{r}_{ij})
\phi_{LM}(\mathbf{R}_k)\right]_{L_TM_T}
\end{eqnarray}
For baryons with only two identical particles, such as $\Lambda$
and $\Sigma$, the spatial wave function has the following form,
\begin{eqnarray}
\Psi_{L_TM_T}(\mathbf{R}, \mathbf{r})=
\left[\phi_{lm}(\mathbf{r}_{ij})
\phi_{LM}(\mathbf{R}_k)\right]_{L_TM_T}
\end{eqnarray}
in which quarks $q_i$ and $q_j$ are identical particles. The
spatial wave function of baryons with three different quarks is
the same as Eq. (19), in which the quark $q_k$ is the heaviest
one.

The relative motion wave functions $\phi_{lm}(\mathbf{r}_{ij})$
and $~\phi_{LM}(\mathbf{R}_k)$ are the superpositions of Gaussian
basis functions with different sizes,
\begin{eqnarray}
\phi_{lm}(\mathbf{r}_{ij})&=&\sum_{n=1}^{n_{max}}c_nN_{nl}r_{ij}^{l}
e^{-\nu_nr_{ij}^2}Y_{lm}(\hat{\mathbf{r}}_{ij})\\
\psi_{LM}(\mathbf{R}_k)&=&\sum_{N=1}^{N_{max}}c_NN_{NL}R_k^{L}
e^{-\nu_{N}R_k^2}Y_{LM}(\hat{\mathbf{R}}_k)
\end{eqnarray}
where $N_{nl}$ and $N_{NL}$ are normalization constants. Gaussian
size parameters $\nu_{n}$ and $\nu_{N}$ are taken as geometric
progression,
\begin{eqnarray}
r_n=r_1a^{n-1},&\nu_{n}=\frac{1}{r^2_n},&
a=\left(\frac{r_{n_{max}}}{r_1}\right)^{\frac{1}{n_{max}-1}}\\
R_N=R_1A^{N-1},&\nu_{N}=\frac{1}{R^2_N},&
A=\left(\frac{R_{N_{max}}}{R_1}\right)^{\frac{1}{N_{max}-1}}
\end{eqnarray}
The numbers $n$ and $l$ ($N$ and $L$) specify, respectively, the
radial and angular momenta excitations with respect to the
Jacobi coordinate $\mathbf{r}$ $(\mathbf{R})$. The angular
momenta $l$ and $L$ are coupled to the total orbit angular
momentum $L_T$.

\begin{figure}[ht] \epsfxsize=2.4in \epsfbox{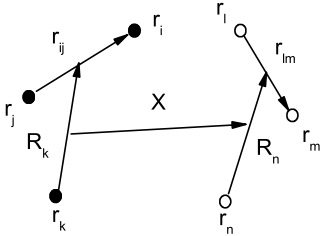}
\caption{Jacob ordinates for a $q^3\bar{q}^3$ system.}
\end{figure}

With regard to a hexaquark system $q^3\bar{q}^3$, the Jacobi
coordinates are shown in Fig.3 and can be expressed as
\begin{eqnarray}
\mathbf{r}_{ij}&=&\mathbf{r}_i-\mathbf{r}_j,~~
\mathbf{R}_k=\mathbf{r}_k-\frac{m_i\mathbf{r}_i+m_j\mathbf{r}_j}{m_i+m_j},\nonumber\\
\mathbf{r}_{lm}&=&\mathbf{r}_l-\mathbf{r}_m,~~
\mathbf{R}_n=\mathbf{r}_n-\frac{m_l\mathbf{r}_l+m_m\mathbf{r}_m}{m_l+m_m},\\
\mathbf{X}&=&\frac{m_i\mathbf{r}_i+m_j\mathbf{r}_j+m_k\mathbf{r}_k}{m_i+m_j+m_k}-
\frac{m_l\mathbf{r}_l+m_m\mathbf{r}_m+m_n\mathbf{r}_n}{m_l+m_m+m_n}\nonumber
\end{eqnarray}
The model wave function with defined quantum numbers $I$ and $J$
can be expressed as,
\begin{eqnarray}
\Psi_{IJ}^{q^3\bar{q}^3}=\sum_{\xi}
c_{\xi}\left[\left[\Phi^{q^3}_{c_1I_1J_1}
\Phi^{\bar{q}^3}_{c_2I_2J_2}\right]_{\xi}F(\mathbf{X})\right]_{IJ}
\end{eqnarray}
$\Phi^{q^3}_{c_1I_1J_1}$ and $\Phi^{\bar{q}^3}_{c_2I_2J_2}$ are
the cluster wave functions of colorful or color singlet baryon
$q^3$ and anti-baryon $\bar{q}^3$, respectively, in which the
spatial functions are same with those of baryons as shown before,
$[\cdots]_\xi$ represents all the needed coupling: color, isospin
and spin. All the possible channels are taken into account in our
multichannel coupling calculation. $F(\mathbf{X})$ is the relative
orbital wave function between $q^3$ and $\bar{q}^3$ clusters, it
also expanded by Gaussians
\begin{eqnarray}
F(\mathbf{X})=\sum_{N^{\prime}=1}^{N^{\prime}_{max}}c_{N^{\prime}}
N_{N^{\prime}L^{\prime}}X^{L^{\prime}}
e^{-\nu_{N^{\prime}}X^2}Y_{L^{\prime}M^{\prime}}(\hat{\mathbf{X}})
\end{eqnarray}
The Gaussian size parameter $\nu_{N^{\prime}}$ is
taken the same form as $\nu_n$ or $\nu_N$.

\section{numerical results and discussions}

The spectrum of the ground states of baryons can be obtained by
solving the three-body Schr\"{o}dinger equation
\begin{eqnarray}
(H_3-E_{IJ})\Phi_{IM_IJM_J}(\mathbf{R},\mathbf{r})=0
\end{eqnarray}
with Rayleigh-Ritz variational principle. The converged results
are arrived by setting $r_1=R_1=0.3$ fm,
$r_{n_{max}}=R_{n_{max}}=2.0$ fm and $n_{max}=N_{max}=5$. The
model parameters are fixed by fitting the experimental data with the
exception that the parameters $\Lambda_0$ and $\mu_0$ are taken
from the paper~\cite{vijande}, $\Lambda_0$=0.187 fm and $\mu_0$=0.113 fm.
The values of model parameters and the masses of the ground baryon
are listed in Table I and Table II, respectively. In general, the model
describe the baryon spectrum well.
\begin{table}[ht]
\caption{Adjustable model parameters.}
\begin{tabular}{ccccccccccccc}\hline\hline
Parameter:  &  ~$m_{ud}$~  &  ~~$m_s$~~  &  ~~$m_c$~~   &  ~~$m_b$~~   &    ~$\alpha_0$     &       $K$        &  $\beta$    \\
\hline
Units:      &      MeV     &    MeV      &     MeV      &     MeV      &        ...         &   MeVfm$^{-2}$   &    fm       \\
Values:     &      313     &    545      &     1800     &     5140     &        5.41        &       400        &   0.47      \\
\hline\hline
\end{tabular}
\end{table}

\begin{table}[ht]
\caption{The masses of the ground states of baryons, unit in MeV,
in which $n$ stands for a $u$- or $d$- quark.}
\begin{tabular}{ccccccccccccc} \hline\hline
Baryons&             ~~~Flavor~~~   &        ~~~$IJ^P$~~~             & ~~Calculated~~   &  Experimental \\
\hline
$N$            &        $nnn$       &    $\frac{1}{2}\frac{1}{2}^+$   &      939         &      939      \\
$\Lambda$      &        $nns$       &    $0\frac{1}{2}^+$             &      1108        &      1116     \\
$\Sigma$       &        $nns$       &    $1\frac{1}{2}^+$             &      1213        &      1195     \\
$\Xi$          &        $nss$       &    $\frac{1}{2}\frac{1}{2}^+$   &      1350        &      1315     \\
$\Delta$       &        $nnn$       &    $\frac{3}{2}\frac{3}{2}^+$   &      1232        &      1232     \\
$\Sigma^*$     &        $nns$       &    $1\frac{3}{2}^+$             &      1382        &      1385     \\
$\Xi^*$        &        $nss$       &    $\frac{1}{2}\frac{3}{2}^+$   &      1528        &      1530     \\
$\Omega^-$     &        $sss$       &    $0\frac{3}{2}^+$             &      1675        &      1672     \\
$\Lambda_c^+$  &        $nnc$       &    $0\frac{1}{2}^+$             &      2287        &      2285     \\
$\Sigma_c$     &        $nnc$       &    $1\frac{1}{2}^+$             &      2480        &      2455     \\
$\Sigma_c^*$   &        $nnc$       &    $1\frac{3}{2}^+$             &      2533        &      2520     \\
$\Xi_c$        &        $nsc$       &    $\frac{1}{2}\frac{1}{2}^+$   &      2620        &      2466     \\
$\Xi_c^*$      &        $nsc$       &    $\frac{1}{2}\frac{3}{2}^+$   &      2670        &      2645     \\
$\Omega_c^0$   &        $ssc$       &    $0\frac{1}{2}^+$             &      2790        &      2695     \\
$\Omega_c^{0*}$&        $ssc$       &    $0\frac{3}{2}^+$             &      2819        &      2766     \\
$\Lambda_b^0$  &        $nnb$       &    $0\frac{1}{2}^+$             &      5600        &      5620     \\
$\Sigma_b$     &        $nnb$       &    $1\frac{1}{2}^+$             &      5816        &      5808     \\
$\Sigma_b^*$   &        $nnb$       &    $1\frac{3}{2}^+$             &      5836        &      5830     \\
$\Xi_b$        &        $nsb$       &    $\frac{1}{2}\frac{1}{2}^+$   &      5948        &      5790     \\
$\Xi_b^*$      &        $nsb$       &    $\frac{1}{2}\frac{3}{2}^+$   &      5966        &      ...     \\
$\Omega_b^-$   &        $ssb$       &    $0\frac{1}{2}^+$             &      6107        &      6071     \\

\hline\hline
\end{tabular}
\end{table}

The color flux-tube model with the model parameters listed in
Table I are used to study the hexaquark systems $q^3\bar{q}^3$. It
should be emphasized that no any new parameter is introduced in
the calculation. The spectrum of hexaquark systems can be obtained
by solving the six-body Schr\"{o}dinger equation
\begin{eqnarray}
(H_6-E_{IJ})\Psi^{q^3\bar{q}^3}_{IJ}=0.
\end{eqnarray}
The converged numerical
results can be obtained by setting $n_{max}$=5, $N_{max}=5$ and
$N^{\prime}_{max}=5$. The minimum and maximum ranges of the bases
are 0.3 fm and 2.0 fm for coordinates $\mathbf{r}$, $\mathbf{R}$
and $\mathbf{X}$, respectively.

\begin{table}[ht]
\caption{The binding energies of the ground states of
hexaquark systems $q^3\bar{q}^3$ with quantum numbers $J^{PC}$,
(unit: MeV), where ``..." means that the corresponding state does
not exist. For the states with lowest energies, all the orbital
angular momenta are set to be zero, therefore the parity of the
systems is negative and the $C$-parity is $(-1)^S$ because of the
baryonia are pure neutral systems.}
\begin{tabular}{ccccccccccccc}\hline\hline
~~~~States~~~                      &   ~~~~~$0^{-+}$~~~~~  &  ~~~~~$1^{--}$~~~~~  &  ~~~~~$2^{-+}$~~~~~ &  ~~~~~$3^{--}$~  \\
\hline
$N\bar{N}$                         &          -46          &         0           &         ...         &         ...       \\
$\Lambda\bar{\Lambda}$             &          -80          &        -46          &          0          &          0        \\
$\Sigma\bar{\Sigma}$               &          -10          &         0           &         ...         &         ..        \\
$\Xi\bar{\Xi}$                     &          -8           &         0           &         ...         &         ...       \\
$\Delta\bar{\Delta}$               &          -80          &         0           &          0          &          0        \\
$\Sigma^*\bar{\Sigma}^*$           &          -20          &         0           &          0          &          0        \\
$\Xi^*\bar{\Xi}^*$                 &          -105         &        -47          &          0          &          0        \\
$\Omega^-\Omega^+$                 &          -8           &         0           &          0          &          0        \\
$\Lambda_c^+\Lambda_c^-$           &          -244         &        -232         &         ...         &         ...       \\
$\Sigma_c\bar{\Sigma}_c$           &          -144         &        -132         &         ...         &         ...       \\
$\Sigma_c^*\bar{\Sigma}_c^*$       &          -34          &         0           &          0          &          0        \\
$\Xi_c\bar{\Xi}_c$                 &          -207         &         0           &         ...         &         ...       \\
$\Xi_c^*\bar{\Xi}_c^*$             &          -168         &        -98          &          0          &          0        \\
$\Omega_c^0\bar{\Omega}_c^0$       &           0           &         0           &         ...         &         ...       \\
$\Omega_c^{0*}\bar{\Omega}_c^{0*}$ &           0           &         0           &          0          &          0        \\
$\Lambda_b^+\Lambda_b^-$           &          -363         &        -360         &         ...         &         ...       \\
$\Sigma_b\bar{\Sigma}_b$           &          -278         &        -277         &         ...         &         ...       \\
$\Sigma_b^*\bar{\Sigma}_b^*$       &          -58          &        -10          &          0          &          0        \\
$\Xi_b\bar{\Xi}_b$                 &          -304         &        -44          &         ...         &         ...       \\
$\Xi_b^*\bar{\Xi}_b^*$             &          -266         &        -200         &         ...         &         ...       \\
$\Omega_b^-\Omega_b^+$             &           0           &         0           &         ...         &         ...       \\

\hline\hline
\end{tabular}
\end{table}

\begin{table}[ht]
\caption{Rms for $\mathbf{r}$, $\mathbf{R}$ and $\mathbf{X}$ of
all possible bound states $q^3\bar{q}^3$ with $J^{PC}=0^{-+}$ and
$1^{--}$, unit in fm.}
\begin{tabular}{c|ccc|ccccccccc}\hline\hline
 ~~~~$J^{PC}$~~~~&                    &   $0^{-+}$ &&& $1^{--}$   \\
\hline Rms&$~\left\langle
{\mathbf{r}^2}\right\rangle^{\frac{1}{2}}$&$\left\langle
{\mathbf{R}^2}\right\rangle^{\frac{1}{2}}$&$\left\langle
\mathbf{X}^2\right\rangle^{\frac{1}{2}}$~~&~~$\left\langle
{\mathbf{r}^2}\right\rangle^{\frac{1}{2}}$&$\left\langle
{\mathbf{R}^2}\right\rangle^{\frac{1}{2}}$&$\left\langle
\mathbf{X}^2\right\rangle^{\frac{1}{2}}$~\\
\hline
$N\bar{N}$                         &   0.81   &    0.71    &  0.68   &  ...      &   ...    &  ...    \\
$\Lambda\bar{\Lambda}$             &   0.76   &    0.66    &  0.49   &  0.77     &   0.69   &   0.54  \\
$\Sigma\bar{\Sigma}$               &   0.86   &    0.67    &  0.87   &  ...      &   ...    &   ...   \\
$\Xi\bar{\Xi}$                     &   0.72   &    0.69    &  0.73   &  ...      &   ...    &   ...   \\
$\Delta\bar{\Delta}$               &   0.93   &    0.82    &  0.62   &  ...      &   ...    &   ...   \\
$\Sigma^*\bar{\Sigma}^*$           &   0.89   &    0.78    &  0.51   &  ...      &   ...    &   ...   \\
$\Xi^*\bar{\Xi}^*$                 &   0.75   &    0.77    &  0.47   &  ...      &   ...    &   ...   \\
$\Omega^-\Omega^+$                 &   0.75   &    0.70    &  0.45   &  ...      &   ...    &   ...   \\
$\Lambda_c^+\Lambda_c^-$           &   0.75   &    0.62    &  0.39   &   0.75    &   0.62   &   0.40  \\
$\Sigma_c\bar{\Sigma}_c$           &   0.85   &    0.65    &  0.42   &   0.85    &   0.66   &   0.41  \\
$\Sigma_c^*\bar{\Sigma}_c^*$       &   0.88   &    0.70    &  0.50   &  ...      &   ...    &   ...   \\
$\Xi_c\bar{\Xi}_c$                 &   0.75   &    0.61    &  0.39   &  ...      &   ...    &   ...   \\
$\Xi_c^*\bar{\Xi}_c^*$             &   0.81   &    0.62    &  0.40   &   0.83    &   0.62   &   0.40  \\
$\Lambda_b^+\Lambda_b^-$           &   0.75   &    0.61    &  0.38   &   0.75    &   0.61   &   0.38  \\
$\Sigma_b\bar{\Sigma}_b$           &   0.83   &    0.64    &  0.38   &   0.83    &   0.64   &   0.38  \\
$\Sigma_b^*\bar{\Sigma}_b^*$       &   0.86   &    0.67    &  0.38   &   0.87    &   0.68   &   0.39  \\
$\Xi_b\bar{\Xi}_b$                 &   0.75   &    0.61    &  0.38   &   0.75    &   0.60   &   0.39  \\
$\Xi_b^*\bar{\Xi}_b^*$             &   0.80   &    0.61    &  0.38   &   0.82    &   0.61   &   0.38  \\
\hline\hline
\end{tabular}
\end{table}

The binding energies, $\Delta E_{J}=E_{IJ}-2M_B$, of the ground
states of the hexaquark systems $q^3\bar{q}^3$ are listed in Table
III. It can be seen that the energies of many low-spin ($J\le 1$)
states lie below the threshold $2M_B$, the ground states are
therefore stable against dissociation into a baryon and an
antibaryon, while they can decay into three mesons. None of
high-spin ($J\ge 2$) states lie below the corresponding threshold.
To check the rationality of the various color structures used,
the spatial configurations of the bound states are calculated by using
the wave functions obtained in solving the
Schr\"{o}dinger equation. The rms for $\mathbf{r}$, $\mathbf{R}$
and $\mathbf{X}$ of all possible bound states $q^3\bar{q}^3$ with
$J^{PC}=0^{-+}$ and $1^{--}$ are listed in Table IV, it can be
seen that they are smaller than 1 fm in our model, so the introducing of
hidden-color configuration is reasonable. The results also show that
the dominant component of the bound states is not a loose
baryon-antibaryon molecule state but a compact hexaquark state, which
is formed by means of the
multibody confinement potential originating from the color
flux-tube picture. Compared with the early baryonia
calculations in the traditional quark models~\cite{PRC73,MPLA26}, where
no non-strange bound state was obtained, the multibody confinement
interaction used in our model can globally give more attraction than
the additive two-body one proportional to the color factor in the
traditional quark models. Furthermore, the anti-confinement
among a color symmetrical quark or antiquark pair does not shown up
in the multibody confinement potential, because no color charge
appeared~\cite{scalar,cyclobutadiene,deng2}.
The similar string model with a multibody confinement potential
was applied to study the stabilities of tetraquark, pentaquark and
hexaquark states, and it was suggested that many compact
multiquark states could exist~\cite{stable}. In addition, the
color-magnetic interaction
$\alpha_s\frac{\mathbf{\sigma}_i\cdot\mathbf{\sigma}_j
\mathbf{\lambda}_i\cdot\mathbf{\lambda}_j}{m_im_j}$ in OGE can
further depress the masses of low-spin states. The color-magnetic
interaction was considered as a binding mechanism and play an
important role in the formation of the famous
H-particle~\cite{hparticle}, which is tentatively below the
threshold of $\Lambda\Lambda$, although it is not confirmed by
experiments so far. However, some researches
of multiquark states indicated that the color-magnetic interaction
as a unique binding mechanism encountered some
difficulties~\cite{difficult}.

From our calculation, a tendency is apparent, the heavier the states,
the deeper the binding and the smaller the size. For example, the binding
energy of $\Lambda\bar{\Lambda}$ is -80 MeV, the distance between two clusters
$\sqrt{\langle \mathbf{X}^2\rangle}$ is 0.49 fm,
while the binding energy of $\Lambda_c\bar{\Lambda}_c$ is -244 MeV
with $\sqrt{\langle \mathbf{X}^2\rangle}$=0.39 fm,
of $\Lambda_b\bar{\Lambda}_b$ is -363 MeV and
$\sqrt{\langle \mathbf{X}^2\rangle}$=0.38 fm. We have similar tendency for
$\Sigma\bar{\Sigma}$-$\Sigma_c\bar{\Sigma}_c$-$\Sigma_b\bar{\Sigma}_b$,
$\Sigma^*\bar{\Sigma}^*$-$\Sigma^*_c\bar{\Sigma}^*_c$-$\Sigma^*_b\bar{\Sigma}^*_b$
and so on. The results agree with the tetraquark state calculations~\cite{ycyang}.

With regard to the light nonstrange hexaquark system
$nnn\bar{n}\bar{n}\bar{n}$, there are two interesting states,
$N\bar{N}$ and $\Delta\bar{\Delta}$ with $J^{PC}=0^{-+}$. The masses
obtained are 1832 MeV and 2384 MeV, respectively, which are
very close to the experimental data of the $X(1835)$ and $X(2370)$
observed in the radiative decay of $J/\psi$ by BES collaboration.
Therefore the bound states $N\bar{N}$ and $\Delta\bar{\Delta}$ may
be the dominant components of the $X(1835)$ and $X(2370)$,
respectively. Our interpretation is consistent with many authors'
points of view in the different theoretical
frameworks~\cite{ppbar-the}. Alternatively, the $X(2370)$ could also
be explained as the bound state $N(1440)\bar{N}$ or
$N\bar{N}(1440)$ with Bethe-Salpeter equation~\cite{zgwang}. For
the light strange hexaquark systems $nns\bar{n}\bar{n}\bar{s}$ and
$nss\bar{n}\bar{s}\bar{s}$, several weakly bounded states,
$\Lambda\bar{\Lambda}$, $\Sigma\bar{\Sigma}$,
$\Sigma^*\bar{\Sigma}^*$, $\Xi\bar{\Xi}$ and $\Xi^*\bar{\Xi}^*$
exist in the color flux-tube model. The mass of the bound state
$\Lambda\bar{\Lambda}$ with $J^{PC}=1^{--}$, 2186 MeV, is close to
the experimental value of the $Y(2175)$. Therefore the dominant
component of $Y(2175)$ could be treated as a bound state
$\Lambda\bar{\Lambda}$ in our model. The interpretation of $Y(2175)$
as the bound state $\Lambda\bar{\Lambda}$ with $J^{PC}=1^{--}$ is
also proposed in other constituent models~\cite{lzhao,abud}.
The weakly bounded state of $\Sigma\bar{\Sigma}$ was also obtained
in Ref.~\cite{abud}. The states $\Lambda\bar{\Lambda}$,
$\Sigma\bar{\Sigma}$ and $\Xi\bar{\Xi}$ were investigated in the
framework of the Bethe-Salpeter equation with a phenomenological
potential and similar conclusions were arrived~\cite{zgwang}.

With respect to the heavy hexaquark systems with a $c\bar{c}$ or
$b\bar{b}$ pair, the states $\Lambda_c^+\Lambda_c^-$,
$\Sigma_c\bar{\Sigma}_c$, $\Xi_c\bar{\Xi}_c$,
$\Xi_c^*\bar{\Xi}_c^*$, $\Lambda_b^+\Lambda_b^-$,
$\Sigma_b\bar{\Sigma}_b$ and $\Xi_b\bar{\Xi}_b$ can form baryonia
with deep binding energies, while the binding energy of states
$\Sigma_c^*\bar{\Sigma}_c^*$ and $\Sigma_b^*\bar{\Sigma}_b^*$ are
several tens of MeVs. Concerning the states $\Omega^-\Omega^+$,
$\Omega_c^0\bar{\Omega}_c^0$, $\Omega_c^{0*}\bar{\Omega}_c^{0*}$
and $\Omega^-_b\Omega^+_b$, only the state $\Omega^-\Omega^+$ has
a shallow binding energy, about 8 MeV, the others are unbound.
Compared with the state $\Omega\bar{\Omega}$, the states
$\Omega_c^0\bar{\Omega}_c^0$, $\Omega_c^{0*}\bar{\Omega}_c^{0*}$
and $\Omega^-_b\Omega^+_b$ have a bigger quark mass, although it
makes the kinetic energies lower, it can also reduce the
color-magnetic interaction, which is disadvantaged to form a bound
state in the model. The heavy baryonia with a
$c\bar{c}$ pair were systematically investigated within the
framework of the one-boson-exchange ($\pi$, $\eta$, $\rho$,
$\omega$, $\phi$ and $\sigma$) model, it is suggested that the
states $\Lambda_c^+\Lambda_c^-$, $\Sigma_c\bar{\Sigma}_c$ and
$\Xi_c\bar{\Xi}_c$ have deep attractive potentials in the
short-distance domain~\cite{nlee}, which is qualitatively
consistent with our conclusions. It seems that the one-boson-exchange
effect in the short-distance can be described by the coupling of
different color flux-tube structures in our model,
which is deserved to be studied in the future work. The
heavy partners $\Lambda_c\bar{\Lambda}_c$ (with mass 4330 MeV) and
$\Lambda_b\bar{\Lambda}_b$ (with mass 10877 MeV)
of the bound states $\Lambda\bar{\Lambda}$ may be used to explain
the states $Y(4260)$ or $Y(4360)$, and $Y_b(10890)$, respectively.
The interpretation is consistent with the point of view in the
paper~\cite{cfqiao,ydchen}.

Those baryon-antibaryon bound states, if they really exist, can be
observed in the corresponding baryon-antibaryon invariant mass
spectrum when they are produced in the $e^+e^-$ annihilation and
charmonium or bottomonium decay processes, they can eventually decay into three
color singlet mesons. Before the occurrence of this decay, the
hidden color hexaquark states must change into several colorless
subsystems by means of the rupture and recombination of color flux
tubes because of a color confinement. This decay mechanism is
similar to compound nucleus formation and therefore should induce
a resonance, which is named as a ``color confined, multi-quark
resonance" state in our models~\cite{resonance}, it is different
from all of those microscopic resonances discussed by S.
Weinberg~\cite{weinberg}.

\section{summary}

The mass spectra of baryon-antibaryon states containing strange,
charm and bottom quarks have
been studied in the color flux-tube model with a multibody
confinement interaction. A powerful numerical method with
high precision, GEM is used in the calculation. The numerical
results indicate
that many low-spin states can not decay into a baryon and an
anti-baryon but into three color singlet mesons by means of the
rupture and recombination of color flux tubes, while the high-spin
states cannot form bound states. The multi-body confinement
interaction as a binding mechanism can globally give more
attractions in the short-distance domain than the two-body one
does in the study of the multiquark calculations, the effect
seems to be equivalent to that of the $\omega$ and $\rho$ meson
exchanges. In addition, the color-magnetic interaction can provide
a further attraction for the low-spin states.

Our predicted bound baryon-antibaryon states $\Sigma\bar{\Sigma}$,
$\Sigma^*\bar{\Sigma}^*$, $\Xi\bar{\Xi}$, $\Xi^*\bar{\Xi}^*$,
$\Sigma_c\bar{\Sigma}_c$, $\Xi_c\bar{\Xi}_c$,
$\Xi_c^*\bar{\Xi}_c^*$, $\Sigma_b\bar{\Sigma}_b$ and
$\Xi_b\bar{\Xi}_b$, if they really exist, can be observed in the
corresponding baryon-antibaryon invariant mass spectrum when they
are produced in the $e^+e^-$ annihilation and charmonium or bottomonium decay
processes. The dominant components of the new hadron states,
$X(1835)$, $X(2370)$, $Y(2175)$, $Y(4260)$ and $Y_b(10890)$, may
be interpreted as $N\bar{N}$, $\Delta\bar{\Delta}$,
$\Lambda\bar{\Lambda}$, $\Lambda_c\bar{\Lambda}_c$ and
$\Lambda_b\bar{\Lambda}_b$ bound states, respectively. The
calculation of the decay properties of the states have to be
invoked to justify the assignment, however, the present difficulty
lies in the lack of the reliable knowledge of the rupture and
recombination of color flux tubes, which is worth being studied in
the future.

\acknowledgments{This research is supported partly by the National
Science Foundation of China under Contract Nos. 11175088,
11035006, 11265017 and Chongqing Natural Science Foundation under
the project No. cstc2013jcyjA00014.}

\end{document}